\input ./style/arxiv-vmsta.cfg
\documentclass[numbers,compress,v1.0.1]{vmsta}
\usepackage{vtexbibtags}

\usepackage{amsfonts}
\usepackage{enumitem}
\usepackage{amsthm}
\usepackage{multirow}

\newtheorem{thm}{Theorem}

\theoremstyle{definition}
\newtheorem{defin}{Definition}

\newtheorem{exmp}{Example}[section]

\def\Cov{\mathbb{C}{\rm ov}}
\def\Corr{\mathbb{C}{\rm orr}}
\def\Var{\mathbb{V}{\rm ar}}

\volume{6}
\issue{2}
\pubyear{2019}
\firstpage{227}
\lastpage{249}
\aid{VMSTA130}
\doi{10.15559/19-VMSTA130}
\articletype{research-article}

\startlocaldefs

\allowdisplaybreaks
\ifdefined\HCode 
\def\index#1{}
\else 
\fi

\endlocaldefs

\begin{document}

\begin{frontmatter}
\pretitle{Research Article}
\title{A copula-based bivariate integer-valued autoregressive process
with application}

\author{\inits{A.}\fnms{Andrius}~\snm{Buteikis}\thanksref{cor1}\ead
[label=e1]{andrius.buteikis@mif.vu.lt}}
\thankstext[type=corresp,id=cor1]{Corresponding author.}
\author{\inits{R.}\fnms{Remigijus}~\snm{Leipus}\ead
[label=e2]{remigijus.leipus@mif.vu.lt}}
\address{Institute of Applied Mathematics,
Faculty of Mathematics and Informatics,
\institution{Vilnius University}, \cny{Lithuania}}



\markboth{A. Buteikis, R. Leipus}{A copula-based bivariate
integer-valued autoregressive process with application}

\begin{abstract}
A bivariate integer-valued autoregressive
process of order 1 (BINAR(1)) with copu\-la-joint innovations is studied.
Different parameter estimation methods are analyzed and compared
via Monte Carlo simulations
with emphasis on estimation of the copula dependence parameter. An empirical
application on defaulted and non-defaulted loan data is carried out using
different combinations of copula functions and marginal distribution
functions covering the cases where both marginal distributions are from the
same family, as well as the case where they are from different distribution
families.\vspace{-4pt}
\end{abstract}

\received{\sday{21} \smonth{8} \syear{2018}}
\revised{\sday{12} \smonth{12} \syear{2018}}
\accepted{\sday{28} \smonth{1} \syear{2019}}
\publishedonline{\sday{12} \smonth{3} \syear{2019}}
\end{frontmatter}

\section{Introduction}
Different financial institutions that issue loans do
this following
company-specific (and/or country-defined) rules which act as a
safeguard 
against loans\index{loans} 
issued to people who are known to be
insolvent. However, 
striving for higher profits might motivate some
companies to issue loans to higher risk clients. Usually 
company's methods for evaluating loan risk are 
not publicly available. However,
one way to 
evaluate if there aren't too many knowingly very high-risk loans
issued, 
and if 
insolvent clients are
adequately separated from responsible clients, would be to look at the quantity of defaulted and
non-defaulted 
loans issued each day. The adequacy of 
company's rules for
issuing loans\index{issuing loans} can be analysed by modelling
via copulas\index{copulas} the dependence between
the number of defaulted loans\index{loans ! defaulted}
and the number of non-defaulted loans\index{loans ! non-defaulted}.
%
The advantage of such approach is that copulas allow to model the
marginal distributions\index{marginal distributions} (possibly from different distribution families)
and their dependence structure (which is described via a copula\index{copulas})
separately. Because of this feature, copulas\index{copulas} were applied to many
different fields, including survival analysis, hydrology, insurance
risk analysis as well as finance (for examples of copula applications,\index{copulas ! applications}
see \cite{Brigo:2010} or \cite{Cherubini:2011}), which also included
the analysis of loans\index{loans} and their default rates.\looseness=1

The dependence 
of the default rate of loans\index{loans} 
on different credit risk categories 
was analysed in \cite{Crook:2011}. 
To model the dependence, copulas\index{copulas} from ten different families were applied
and three model selection tests were carried out. Because of the small
sample size (24 observations per risk category) most of the copula
families\index{copulas ! families} were not rejected and a single best copula model\index{copulas ! model} was not
selected. To analyse whether dependence is affected by time, Fenech et
al.\ \cite{Fenech:2015} estimated the dependence 
among four different
loan default indexes before the global financial crisis and after.
They have found that the dependence was different in these periods. Four copula families\index{copulas ! families} were used to estimate the dependence between the
default index pairs.
While these studies were carried out for continuous data,
discrete models created with copulas\index{copulas} are less investigated:
Genest and Ne\v{s}lehov\'a
\cite{Genest:2007} discussed the differences and challenges of using
copulas\index{copulas} for discrete data compared to continuous data. Note that
the previously mentioned studies assumed that the data does not depend
on its own previous values. By using bivariate integer-valued autoregressive
models (BINAR\index{BINAR}) it is possible to account for both the discreteness and
autocorrelation of the data. Furthermore, copulas\index{copulas} can be used to model
the dependence of innovations\index{innovations} in the BINAR\index{BINAR}(1) models: Karlis and Pedeli
\cite{Karlis:2013} used the Frank copula\index{Frank copula} and the normal copula to model the
dependence of the innovations\index{innovations} of the BINAR\index{BINAR}(1) model.

In this paper we expand on using copulas in BINAR models\index{copulas ! in BINAR models} by analysing
additional copula families\index{copulas ! families} for the innovations\index{innovations} of the BINAR\index{BINAR}(1) model
and analyse different methods for BINAR(1) model parameter estimation. 
We also present a two-step method for
the parameter estimation of the
BINAR\index{BINAR}(1) model, where we estimate the model parameters separately from
the dependence parameter of the copula.\index{copulas} These estimation methods
(including the one used in \cite{Karlis:2013}) are compared via Monte
Carlo simulations. Finally, in order to analyse the presence of
autocorrelation and copula dependence in loan data,\index{copulas ! dependence in loan data} an empirical
application is carried out for empirical weekly loan data.

The paper is organized as follows. Section~\ref{CH:BINAR} presents the
BINAR\index{BINAR}(1) process and its main properties, Section \ref{CH:COPULA}
presents the main properties of copulas\index{copulas} as well as some copula
functions.\index{copulas ! functions} Section~\ref{CH:ESTMATION} compares different estimation
methods for the BINAR\index{BINAR}(1) model and the dependence parameter of copulas\index{copulas}
via Monte Carlo simulations. In Section~\ref{CH:APPLICATION} an
empirical application is carried out using different combinations of
copula functions\index{copulas ! functions} and marginal distribution functions.\index{marginal distribution functions} Conclusions are
presented in Section~\ref{CH:CONCLUSION}.\looseness=1

\section{The bivariate INAR\index{bivariate INAR}(1) process}\label{CH:BINAR}

The BINAR\index{BINAR}(1) process was introduced in \cite{Pedeli:2011}. In this
section we will provide the definition of the BINAR\index{BINAR}(1) model and will
formulate its properties.

\begin{defin}\label{DEF:BINAR}
Let $\mathbf{R}_t=[R_{1,t},R_{2,t}]'$, $t \in\mathbb{Z}$, be a sequence
of independent identically distributed (i.i.d.) nonnegative
integer-valued bivariate random variables. A bivariate integer-valued
autoregressive process of order 1 (BINAR\index{BINAR}(1)), $\mathbf
{X}_t=[X_{1,t},X_{2,t}]'$, $t\in\mathbb{Z}$, is defined as:
%
\begin{equation}
\label{eq:BIN} \mathbf{X}_t = \mathbf{A}\circ
\mathbf{X}_{t-1}+\mathbf{R}_t = %
\begin{bmatrix}
\alpha_1 & 0 \\
0 & \alpha_2
\end{bmatrix} %
\circ %
\begin{bmatrix}
X_{1,t-1} \\
X_{2,t-1}
\end{bmatrix}
+ %
\begin{bmatrix}
R_{1,t} \\
R_{2,t}
\end{bmatrix} %
, \quad t \in\mathbb{Z},
\end{equation}
where $\alpha_j\in[0,1)$, $j=1,2$, and the symbol `$\circ$' is the
thinning operator\index{thinning operator} which also acts as the matrix multiplication. So the
$j$th ($j=1,2$) element is defined as an INAR process\index{INAR process} of order 1 (INAR\index{INAR}(1)):
%
\begin{equation}
\label{eq:INAR} X_{j,t} = \alpha_j \circ
X_{j,t-1} + R_{j,t}, \quad t \in\mathbb{Z},
\end{equation}
where $\alpha_j \circ X_{j,t-1} := \sum_{i=1}^{X_{j,t-1}} Y_{j,t,i}$
and $Y_{j,t,1}, Y_{j,t,2},\dots$ is a sequence of i.i.d.\ Bernoulli
random variables with
$\mathbb{P}(Y_{j,t, i} = 1) = \alpha_j = 1-\mathbb{P}(Y_{j,t, i} = 0)$,
$\alpha_j \in[0,1)$, such that these sequences are mutually
independent and independent of the sequence $\mathbf{R}_t$, $t\in
\mathbb{Z}$. For each $t$, $\mathbf{R}_t$ is independent of $\mathbf
{X}_s$, $s<t$.
\end{defin}

Properties of the thinning operator\index{thinning operator}
are provided in \cite
{Pedeli:2011:PhD} and \cite{Silva:2005} with proofs for selected few.
We present the main properties of the thinning operator\index{thinning operator} which will be
used later on in the case of BINAR\index{BINAR}(1) model. Denote by `$\stackrel
{d}{=}$' the equality of distributions.

%
\begin{thm}[Thinning operator\index{thinning operator} properties]\label{TH:ThinningOp}
Let $X, X_1, X_2$ be nonnegative integer-valued random variables, such
that $\mathbb{E}Z^2 < \infty$, $Z \in\{ X, X_1, X_2 \}$, $\alpha,\alpha
_1, \alpha_2 \in[0,1)$ and let `$\circ$' be the thinning operator.\index{thinning operator}
Then the following properties hold:
\begin{enumerate}[label=(\alph*)]
\item$\alpha_1 \circ(\alpha_2 \circ X) \stackrel{d}{=} (\alpha_1
\alpha_2) \circ X$;
\item$\alpha\circ(X_1 + X_2) \stackrel{d}{=} \alpha\circ X_1 +
\alpha\circ X_2$;
\item$\mathbb{E}(\alpha\circ X) = \alpha\mathbb{E}(X)$;
\item$\Var(\alpha\circ X) = \alpha^2 \Var(X) + \alpha(1-\alpha)\mathbb{E}(X)$;
\item$\mathbb{E}((\alpha\circ X_1)X_2) = \alpha\mathbb{E}(X_1X_2)$;
\item$\Cov(\alpha\circ X_1, X_2) = \alpha\Cov(X_1,X_2)$;
\item$\mathbb{E}((\alpha_1 \circ X_1)(\alpha_2 \circ X_2)) = \alpha_1
\alpha_2 \mathbb{E}(X_1X_2)$.
\end{enumerate}
\end{thm}
$X_{j,t}$, defined in eq.\ \eqref{eq:INAR}, has two random components:
the survivors of the elements of the process at time $t-1$, each with
the probability of survival $\alpha_j$, which are denoted by $\alpha_j
\circ X_{j,t-1}$, and the elements which enter in the system
in the interval $(t-1,t]$, which are called arrival elements and
denoted by $R_{j,t}$. We can obtain a moving average representation by
substitutions and the properties of the thinning operator\index{thinning operator} as in \cite
{Al-Osh_Alzaid:1987} or \cite[p.\ 180]{Fokianos:2002}:
%
\begin{align}
\label{eq:infBIN} X_{j,t} &= \alpha_j \circ
X_{j,t-1} + R_{j,t} \stackrel{d} {=} \sum
_{k=0}^\infty\alpha^k_j
\circ R_{j,t-k}, \quad j = 1,2, t\in \mathbb{Z},
\end{align}
where convergence on the right-hand side holds a.s.

Now we present some properties of the BINAR\index{BINAR}(1) model. They will be used
when analysing some of 
parameter estimation methods.
The proofs for these properties can be easily derived and
some of them are provided in \cite{Pedeli:2011:PhD}.
%
\begin{thm}[Properties of the BINAR\index{BINAR}(1) process]
Let {\upshape
$\textbf{X}_t = (X_{1,t}, X_{2,t})'$} be a nonnegative integer-valued
time series given in Def.\ \ref{DEF:BINAR} and $\alpha_j \in[0,1)$, $j
= 1,2$. Let {\upshape$\textbf{R}_t = (R_{1,t}, R_{2,t})'$}, $t\in
\mathbb{Z}$, be nonnegative integer-valued random
variables with $\mathbb{E}(R_{j,t}) = \lambda_j$ and $\Var(R_{j,t}) =
\sigma^2_j < \infty$, $j = 1,2$. \label{TH:BINAR_PROPERTIES}
Then the following properties hold:
\begin{enumerate}[label=(\alph*)]
\item$\mathbb{E}X_{j,t} = \mu_{X_j} = \frac{\lambda_j}{1 - \alpha_j}$;
\item$\mathbb{E}(X_{j,t}|X_{j,t-1}) = \alpha_j X_{j,t-1} + \lambda_j$;
\item$\Var(X_{j,t}) = \sigma^2_{X_j} = \dfrac{\sigma^2_j + \alpha_j
\lambda_j}{1-\alpha_j^2}$;
\item$\Cov(X_{i,t}, R_{j,t}) = \Cov(R_{i,t}, R_{j,t})$, $i \neq j$;
\item$\Cov(X_{j,t}, X_{j,t+h}) = \alpha_j^h \sigma^2_{X_j},\ h\ge0$;
\item$\Corr(X_{j,t}, X_{j,t+h}) = \alpha^h_j$, $h \geq0$;
\item$
\begin{aligned}[t] \label{BINAR:PROERTY_COV}
\Cov(X_{i,t}, X_{j,t+h}) = \dfrac{\alpha^h_j}{1-\alpha_i\alpha_j}\,\Cov
(R_{i,t},R_{j,t})
\end{aligned}
$, $i \neq j$, $h \geq0$;
\item$
\begin{aligned}[t] \label{BINAR:PROERTY_CORR}
\Corr(X_{i,t+h}, X_{j,t}) = \dfrac{\alpha^h_i \sqrt{(1-\alpha
^2_i)(1-\alpha^2_j})}{(1-\alpha_i\alpha_j)\sqrt{(\sigma^2_i+\alpha
_i\lambda_i)(\sigma^2_j+\alpha_j\lambda_j)}}\,\Cov(R_{i,t},R_{j,t})
\end{aligned}
$,\break$i \neq j$, $h \geq0$;
\end{enumerate}
\end{thm}
Similarly to \eqref{eq:infBIN}, 
we have that
\begin{align*}
\mathbf{X}_t \stackrel{d} {=} \sum
_{k=0}^\infty\mathbf{A}^k \circ
\mathbf{R}_{t-k},
\end{align*}
where convergence on the right-hand side holds a.s.

Hence, the distributional properties of the BINAR\index{BINAR}(1) process can be
studied in terms of $\textbf{R}_t$ values.
Note also, that according to \cite{Latour:1997}, if $\alpha_j\in
[0,1)$, $j=1,2$, then there exists a unique stationary nonnegative
integer-valued sequence
$\mathbf{X}_t$, $t \in\mathbb{Z}$, satisfying \eqref{eq:BIN}.

From the covariance and correlation
(see \ref{BINAR:PROERTY_COV} and \ref{BINAR:PROERTY_CORR} in
Theorem \ref{TH:BINAR_PROPERTIES}) of the BINAR\index{BINAR}(1) process we see that the dependence between
$X_{1,t}$ and $X_{2,t}$ depends on the joint distribution of the
innovations\index{innovations} $R_{1,t}$, $R_{2,t}$. Pedeli and Karlis
\cite{Pedeli:2011} analysed BINAR(1) models when the innovations\index{innovations} were
linked by either a bivariate Poisson or a bivariate negative binomial
distribution,\index{negative binomial distribution} where the covariance of the innovations\index{innovations} can be easily
expressed in terms of their joint distribution parameters. Karlis and
Pedeli \cite{Karlis:2013} analysed two cases when the distributions of
innovations\index{innovations} of a BINAR\index{BINAR}(1) model are linked by either the Frank copula\index{Frank copula} or
a normal copula with either Poisson or negative binomial marginal
distributions.\index{negative binomial marginal distribution} We will expand 
their work by analysing additional
copulas\index{copulas} for the BINAR\index{BINAR}(1) model innovation distribution as well as
estimation methods for the distribution parameters.

\section{Copulas\index{copulas}}\label{CH:COPULA}

In this section we recall the definition and main properties of
bivariate copulas,\index{bivariate copulas} mainly following
\cite{Genest:2007,Nelsen:2006} and \cite{Trivedi:2007} for the
continuous and discrete settings.

\subsection{Copula definition\index{copulas ! definition} and properties}

Copulas\index{copulas} are used for modelling the dependence between several random
variables. The main advantage of
using copulas\index{copulas} is that they allow to model the marginal distributions\index{marginal distributions}
separately from their joint
distribution. In this paper we are using two-dimensional copulas which
are defined as follows:
\begin{defin}\label{DEF:COPULA}
A $2$-dimensional copula $C:[0,1]^2 \rightarrow[0,1]$ is a function
with the following properties:
\begin{enumerate}
\item[(i)] for every $u,v \in[0,1]$:
%
\begin{align}
C(u,0) = C(0,v) = 0;
\end{align}
\item[(ii)] for every $u, v \in[0,1]$:
%
\begin{align}
C(u,1) = u, \quad C(1,v) = v;
\end{align}
\item[(iii)] for any $u_1, u_2, v_1, v_2 \in[0,1]$ such that $u_1 \leq
u_2$ and $v_1 \leq v_2$:
%
\begin{align}
C(u_2, v_2) - C(u_2,
v_1) - C(u_1, v_2) +
C(u_1, v_1) \geq0
\end{align}
(this is also called \textit{the rectangle inequality}).
\end{enumerate}

The theoretical foundation of copulas\index{copulas} is given by Sklar's theorem:
\end{defin}
%
%
\begin{thm}[\cite{Sklar:1959}]\label{TH:Sklar} Let $H$ be a joint
cumulative distribution function (cdf) with mar\-ginal distributions\index{marginal distributions}
$F_1,F_2$. Then there exists a copula $C$ such that for all $(x_1,x_2)
\in\break[-\infty, \infty]^2$:
%
\begin{align}
\label{Sklarth} H(x_1,x_2) = C
\bigl(F_1(x_1),F_2(x_2)
\bigr).
\end{align}
If $F_i$ is continuous for $i=1,2$ then $C$ is unique; otherwise $C$ is
uniquely determined only on ${\rm Ran} (F_1) \times{\rm Ran} (F_2)$,
where ${\rm Ran} (F)$ denotes the range of the cdf $F$. Conversely, if
$C$ is a copula\index{copulas} and $F_1,F_2$ are distribution functions, then the
function $H$, defined by equation \eqref{Sklarth} is a joint cdf with
marginal distributions\index{marginal distributions} $F_1,F_2$.
\end{thm}

If a pair of random variables $(X_1, X_2)$ has continuous marginal cdfs
$F_i(x), i = 1,2$, then by applying the probability integral
transformation one can transform them into random variables $(U_1,U_2)
= (F_1(X_1),F_2(X_2))$ with uniformly distributed mar\-ginals which can
then be used when modelling their dependence via a copula.\index{copulas} More about
Copula theory,\index{copulas ! theory} properties and applications can be found in \cite
{Nelsen:2006} and \cite{Joe:2015}.

\subsection{Copulas\index{copulas} with discrete marginal distributions\index{marginal distributions}}
Since innovations\index{innovations} of a BINAR\index{BINAR}(1) model are nonnegative integer-valued
random variables, one needs to consider
copulas linking\index{copulas ! linking} discrete distributions. In this section we will mention
some of the key differences when copula
marginals\index{copulas ! marginals} are discrete rather than continuous.

Firstly, as mentioned in Theorem~\ref{TH:Sklar}, if $F_1$ and $F_2$ are
discrete marginals then a unique copula representation
exists only for values in the range of ${\rm Ran}(F_1)\times{\rm
Ran}(F_2)$. However, the lack of uniqueness does not pose
a problem in empirical applications because it implies that there may
exist more than one copula\index{copulas} which describes the distribution
of the empirical data.
Secondly, regarding concordance and discordance, the discrete case has
to allow for ties (i.e.\ when two variables have the same value), so
the concordance measures (Spearman's rho and Kendal's tau) are
margin-dependent, see \cite{Trivedi:2007}. There are several
modifications proposed for Spearman's rho, however, none of them are
margin-free. Furthermore,
Genest and Ne\v{s}lehov\'a \cite{Genest:2007}
state that estimators of
the dependence parameter $\theta$ based on Kendall's tau or its
modified versions are biased, and estimation techniques based on maximum
likelihood are recommended. As such, we will not examine estimation
methods based on concordance measures. Another difference from the
continuous case is the use of the probability mass function (pmf)
instead of the probability density function when estimating the model
parameters which will be seen in Section~\ref{CH:ESTMATION}.

\subsection{Some concrete copulas\index{copulas}} \label{ch:kop}
In this section we will present several bivariate copulas,\index{bivariate copulas} which will
be used later when constructing and evaluating
the BINAR\index{BINAR}(1) model. For all the copulas discussed, the following
notation is used: $u_1 := F_1(x_1)$, $u_2:=F_2(x_2)$,
where $F_1,F_2$ are marginal cumulative distribution functions (cdfs) of discrete random variables, and
$\theta$ is the dependence parameter.

\subsubsection*{Farlie--Gumbel--Morgenstern copula}
\noindent The Farlie--Gumbel--Morgenstern (FGM) copula\index{copulas} has the following form:
%
\begin{align}
\label{EQ:FGM} C(u_1, u_2; \theta) &=
u_1 u_2\bigl(1+\theta(1-u_1)
(1-u_2)\bigr).
\end{align}
The dependence parameter $\theta$ can take values from the interval
$[-1,1]$. If $\theta=0$,
then the FGM copula\index{FGM copula} collapses to independence. Note that the FGM copula\index{FGM copula} can only model weak dependence
between two marginals
(see \cite{Nelsen:2006}). The copula\index{copulas} when $\theta= 0$ is called a
product (or independence) copula:\index{copulas}
%
\begin{align}
C(u_1, u_2) &= u_1u_2.
\end{align}
Since the product copula\index{product copula} corresponds to independence, it is important
as a benchmark.

\subsubsection*{Frank copula\index{Frank copula}}
\noindent
The Frank copula\index{Frank copula} has the following form:
\begin{align*}
C(u_1, u_2; \theta) &= -\dfrac{1}{\theta}\log
\biggl( 1 + \dfrac{(\exp
(-\theta u_1)-1)(\exp(-\theta u_2)-1)}{\exp(-\theta)-1} \biggr).
\end{align*}
The dependence parameter $\theta$ can take values from $(-\infty, \infty
)\setminus\{0\}$. The Frank copula\index{Frank copula} allows for both positive and negative
dependence between the marginals.

\subsubsection*{Clayton copula\index{Clayton copula}}
\noindent The Clayton copula\index{Clayton copula} has the following form:
%
\begin{align}
\label{COP:CLAYTON} C(u_1, u_2; \theta) &= \max\bigl
\{u_1^{-\theta} + u_2^{-\theta} - 1 , 0
\bigr\} ^{-\frac{1}{\theta}},
\end{align}
with the dependence parameter $\theta\in[-1, \infty) \setminus\{0\}
$. The marginals become independent when $\theta\rightarrow0$.
It can be used when the correlation between two random variables
exhibits a strong left tail dependence -- if smaller values are
strongly correlated and hight values are less correlated. The Clayton
copula\index{Clayton copula} can also account for negative dependence when $\theta\in[-1,0)$.
For more properties of this copula,\index{copulas} see the recent paper by Manstavi\v
{c}ius and Leipus \cite{Manstavicius_Leipus:2017}.

\section{Parameter estimation of the copula-based BINAR(1) model}\label
{CH:ESTMATION}
In this section we examine different BINAR(1) model parameter
estimation methods and provide a two-step 
method for separate estimation of the copula
dependence parameter.\index{copulas ! dependence parameter} 
Estimation methods are compared via
Monte Carlo simulations.
\noindent Let $\textbf{X}_t = (X_{1,t}, X_{2,t})'$ be a non-negative
integer-valued time series given in Def.\ \ref{DEF:BINAR},
where the joint distribution of $(R_{1,t}, R_{2,t})'$, with marginals
$F_1,F_2$, is linked by a copula $C(\cdot, \cdot)$:
\begin{align*}
\mathbb{P}(R_{1,t}\le x_1, R_{2,t}\le
x_2) &= C\bigl(F_1(x_1),
F_2(x_2)\bigr)
\end{align*}
and let $C(u_1,u_2)=C(u_1,u_2;\theta)$, where $\theta$ is a dependence
parameter.

\subsection{Conditional least squares estimation}
\noindent
The Conditional least squares\index{Conditional least squares (CLS)} (CLS) estimator minimizes the squared
distance between $\textbf{X}_t$ and its conditional expectation.
Similarly to the method in \cite{Silva:2005} for the INAR\index{INAR}(1) model, we
construct the CLS estimator\index{CLS estimators} in the case of the BINAR\index{BINAR}(1) model.

Using Theorem \ref{TH:ThinningOp} we can write the vector of
conditional means as
%
\begin{align}
\boldsymbol{\mu}_{t|t-1} := %
\begin{bmatrix}
\mathbb{E}(X_{1,t}|X_{1,t-1}) \\
\mathbb{E}(X_{2,t}|X_{2,t-1})
\end{bmatrix}
= %
\begin{bmatrix}
\alpha_1 X_{1,t-1} + \lambda_1 \\
\alpha_2 X_{2,t-1} + \lambda_2
\end{bmatrix} %
,
\end{align}
where $\lambda_j\,{:=}\,\mathbb{E}R_{j,t}$, $j\,{=}\,1,2$.
In order to calculate the CLS estimators\index{CLS estimators} of $(\alpha_1, \alpha_2,
\lambda_1, \lambda_2)$ we define the vector of residuals as the
difference between the observations and their conditional expectation:
%
\begin{align}
\nonumber
\textbf{X}_t - \boldsymbol{\mu}_{t|t-1} &=
\begin{bmatrix}
X_{1,t} - \alpha_1 X_{1,t-1} - \lambda_1 \\
X_{2,t} - \alpha_2 X_{2,t-1} - \lambda_2
\end{bmatrix} %
.
\end{align}
Then, given a sample of $N$ observations, $\textbf{X}_1,\dots, \textbf
{X}_N$, the
CLS estimators\index{CLS estimators} of $\alpha_j, \lambda_j$, $j = 1,2$, are found by
minimizing the sum
\begin{align*}
Q_j(\alpha_j,\lambda_j) &:= \sum
_{t=2}^N (X_{j,t} -
\alpha_j X_{j,t-1} - \lambda_j)^2
\ \longrightarrow\ \min_{\alpha_j, \lambda_j}, \quad j = 1,2.
\end{align*}
By taking the derivatives with respect to $\alpha_j$ and $\lambda_j$,
$j = 1,2$, and equating them to zero we get:
%
\begin{align}
\label{EQ:CLS_ALPHA} \hat{\alpha}^{\rm CLS}_j =
\frac{\sum_{t=2}^N (X_{j,t} -\bar
{X}_j)(X_{j,t-1} - \bar{X}_j)}{\sum_{t=2}^N (X_{j,t-1} - \bar{X}_j)^2}
\end{align}
and
%
\begin{align}
\label{EQ:CLS_LAMBDA} \hat{\lambda}^{\rm CLS}_j &=
\frac{1}{N-1} \Biggl(\sum_{t=2}^N
X_{j,t} - \hat{\alpha}^{\rm CLS}_j \sum
_{t=2}^N X_{j,t-1} \Biggr).
\end{align}
The asymptotic properties of the CLS estimators\index{CLS estimators} for the INAR\index{INAR}(1) model
case are provided in \cite{Latour:1998,Silva:2005,Barczy:2010} and can be applied to the BINAR\index{BINAR}(1) parameter estimates,
specified via equations \eqref{EQ:CLS_ALPHA} and \eqref{EQ:CLS_LAMBDA}.
By the fact that the $j$-th component of the BINAR\index{BINAR}(1) process is an
INAR\index{INAR}(1) itself, we can formulate the following theorem for the marginal
parameter vector distributions (see \cite{Barczy:2010}):
%
\begin{thm} 
Let $\mathbf{X}_t =  ( X_{1,t}, X_{2,t}  )'$ be defined in
Def.\ \ref{DEF:BINAR} and let the parameter vector of \eqref{eq:INAR}
be $(\alpha_j, \lambda_j)'$. Assume that $\widehat{\alpha}^{\rm CLS}_j$
and $\widehat{\lambda}^{\rm CLS}_j$ are the {\rm CLS} estimators\index{CLS estimators} of
$\alpha_j$ and $\lambda_j$, $j = 1,2$. Then:
\begin{align*}
\sqrt{N} %
\begin{pmatrix}
\widehat{\alpha}^{\rm CLS}_j - \alpha_j \\[3pt]
\widehat{\lambda}^{\rm CLS}_j - \lambda_j
\end{pmatrix} %
\stackrel{d} {
\longrightarrow} \mathcal{N} ( \boldsymbol{0}_2,
\mathbf{B}_j ),
\end{align*}
where
\begin{align*}
\mathbf{B}_j &= %
\begin{bmatrix}
\mathbb{E} X_{j,t}^2 & \mathbb{E} X_{j,t} \\[3pt]
\mathbb{E} X_{j,t} & 1
\end{bmatrix}
^{-1} \mathbf{A}_j %
\begin{bmatrix}
\mathbb{E} X_{j,t}^2 & \mathbb{E} X_{j,t} \\[3pt]
\mathbb{E} X_{j,t} & 1
\end{bmatrix} %
^{-1},
\\
\mathbf{A}_j &= \alpha_j(1-\alpha_j)
\begin{bmatrix}
\mathbb{E} X_{j,t}^3 & \mathbb{E} X_{j,t}^2 \\[3pt]
\mathbb{E} X_{j,t}^2 & \mathbb{E} X_{j,t}
\end{bmatrix} %
+ \sigma^2_j
\begin{bmatrix}
\mathbb{E} X_{j,t}^2 & \mathbb{E} X_{j,t} \\[3pt]
\mathbb{E} X_{j,t} & 1
\end{bmatrix} %
,\quad j=1,2.
\end{align*}
Here, according to {\rm BINAR\index{BINAR}(1)} properties in Theorem \ref
{TH:BINAR_PROPERTIES},
\begin{align*}
\mathbb{E} X_{j,t} ={}& \frac{\lambda_j}{1-\alpha_j}, \ \ \mathbb{E}
X_{j,t}^2 = \frac{\sigma_j^2+\alpha_j\lambda_j}{1-\alpha^2_j}+\frac
{\lambda^2_j}{(1-\alpha_j)^2},
\\
\mathbb{E} X_{j,t}^3 ={}& \dfrac{\mathbb{E} R_{j,t}^3 - 3 \sigma_j^2(1 +
\lambda_j) - \lambda_j^3 + 2 \lambda_j}{1-\alpha_j^3} + 3
\dfrac{\sigma
^2_j + \alpha_j \lambda_j}{1 - \alpha_j^2} - 2 \dfrac{\lambda_j}{1 -
\alpha_j}
\\
&{}  + 3 \dfrac{\lambda_j(\sigma^2_j + \alpha_j \lambda_j)}{(1-\alpha
_j)(1-\alpha_j^2)} + \dfrac{\lambda_j^3}{(1 - \alpha_j)^3}.
\end{align*}
\end{thm}
For the Poisson marginal distribution\index{marginal distributions} case the asymptotic variance
matrix can be expressed as (see \cite{Freeland:2005})
\begin{align*}
\mathbf{B}_j = %
\begin{bmatrix}
\dfrac{\alpha_j(1-\alpha_j)^2}{\lambda_j} + 1 - \alpha^2_j &
-(1+\alpha_j)\lambda_j\\
-(1+\alpha_j)\lambda_j & \lambda_j + \dfrac{1+\alpha_j}{1-\alpha
_j}\lambda_j^2
\end{bmatrix} %
,\quad j = 1,2.
\end{align*}
Furthermore, for a more general case, \cite{Latour:1997} proved that
the CLS estimators\index{CLS estimators} of a multivariate generalized integer-valued
autoregressive process (GINAR\index{GINAR}) are asymptotically normally distributed.

Note that
%
\begin{align}
\label{EQ:COV_CLS} \mathbb{E}(X_{1,t} - \alpha_1
X_{1,t-1} - \lambda_1) (X_{2,t} -
\alpha_2 X_{2,t-1} - \lambda_2) &=
\Cov(R_{1,t}, R_{2,t}),
\end{align}
which follows from
\begin{align*}
&\mathbb{E}(X_{1,t} - \alpha_1 X_{1,t-1} -
\lambda_1) (X_{2,t} - \alpha _2
X_{2,t-1} - \lambda_2)
\\
&\quad = \mathbb{E} (\alpha_1 \circ X_{1,t-1} -
\alpha_1 X_{1,t-1}) (\alpha_2 \circ
X_{2,t-1} - \alpha_2 X_{2,t-1})
\\
&\qquad + \mathbb{E} (\alpha_1 \circ X_{1,t-1} -
\alpha_1 X_{1,t-1}) (R_{2,t}-
\lambda_2)
\\
&\qquad + \mathbb{E}(\alpha_2 \circ X_{2,t-1} -
\alpha_2 X_{2,t-1}) (R_{1,t}-
\lambda_1)
\\
&\qquad + \mathbb{E}(R_{1,t}- \lambda_1) (R_{2,t}-
\lambda_2)
\end{align*}
since the first three summands are zeros.

%
\begin{exmp} Assume that the joint pmf of $(R_{1,t},
R_{2,t})$ is given by bivariate
Poisson distribution:
\begin{align*}
\mathbb{P}(R_{1,t}=k, R_{2,t}=l) &= \sum
_{i=0}^{\min\{k,l\}} \frac
{(\lambda_1-\lambda)^{k-i}(\lambda_2-\lambda)^{l-i} \lambda
^i}{(k-i)!(l-i)! i!}\, {\rm
e}^{-(\lambda_1+\lambda_2-\lambda)},
\end{align*}
where $k,l = 0, 1,...$, $\lambda_j>0$, $j=1,2$, $0\le\lambda<\min\{\lambda_1,\lambda_2\}
$. Then, for each ${j=1,2}$, the marginal
distribution\index{marginal distributions} of $R_{j,t}$ is Poisson with parameter $\lambda_j$ and
\\$\Cov(R_{1,t}, R_{2,t})=\lambda$. If $\lambda= 0$
then the two variables are independent.
\end{exmp}

%
\begin{exmp} Assume that the joint pmf of $(R_{1,t},
R_{2,t})$ is bivariate negative binomial distribution\index{negative binomial distribution}
given by
\begin{align*}
\mathbb{P}(R_{1,t}=k, R_{2,t}=l) ={}& \frac{\varGamma(\beta+k+l)}{\varGamma
(\beta)k!l!}
\biggl(\frac{\lambda_1}{\lambda_1+\lambda_2+\beta} \biggr)^k \biggl(\frac{\lambda_2}{\lambda_1+\lambda_2+\beta}
\biggr)^l\\
&{}\times \biggl(\frac{\beta}{\lambda_1+\lambda_2+\beta} \biggr)^\beta,
\end{align*}
where $k,l = 0, 1,...$, $\lambda_j>0$, $j=1,2$, $\beta>0$. Then, for each $j=1,2$, the marginal
distribution\index{marginal distributions} of $R_{j,t}$ is negative binomial\index{negative binomial} with parameters $\beta$
and $p_j=\beta/(\lambda_j+\beta)$
and $\mathbb{E} R_{j,t}=\lambda_j$, $\Var(R_{j,t})=\lambda_j (1+\beta
^{-1}\lambda_j)$, $\Cov(R_{1,t}, R_{2,t})=\beta^{-1} \lambda_1\lambda
_2$. Thus,
bivariate negative binomial distribution\index{negative binomial distribution} is more flexible than
bivariate Poisson due to overdispersion parameter $\beta$.
\end{exmp}

Assume now that the Poisson innovations\index{Poisson innovations} $R_{1,t}$ and $R_{2,t}$ with
parameters $\lambda_1$ and $\lambda_2$, respectively, are
linked by a
copula\index{copulas} with
the dependence parameter $\theta$. Taking into account equality \eqref{EQ:COV_CLS},
we can estimate $\theta$ by minimizing the sum of squared differences
%
\begin{align}
\label{EQ:CLS_COP} S &= \sum_{t=2}^N
\bigl(R^{\rm CLS}_{1,t}R^{\rm CLS}_{2,t} -
\gamma\bigl(\hat \lambda^{\rm CLS}_1,\hat
\lambda_2^{\rm CLS};\theta\bigr) \bigr)^2,
\end{align}
where
\begin{align*}
R^{\rm CLS}_{j,t} &:= X_{j,t} - \hat{
\alpha}^{\rm CLS}_j X_{j,t-1} - \hat{
\lambda}^{\rm CLS}_j, \quad j = 1,2,
\\
\gamma(\lambda_1,\lambda_2;\theta)&:=
\Cov(R_{1,t},R_{2,t}) \ = \sum_{k,l=1}^\infty
k l\, c\bigl(F_1(k;\lambda _1),
F_2(l;\lambda_2);\theta\bigr) - \lambda_1
\lambda_2.
\end{align*}
Here, $c(F_1(k;\lambda_1), F_2(s;\lambda_2);\theta)$ is the joint pmf:
%
\begin{align}
c\bigl(F_1(k;\lambda_1), F_2(l;
\lambda_2);\theta\bigr) ={}& \mathbb{P}(R_{1,t} = k,
R_{2,t} = l )
\nonumber
\\
={}& C\bigl(F_1(k;\lambda_1),F_2(s;
\lambda_2);\theta\bigr)
\nonumber
\\
&{} - C\bigl(F_1(k - 1;\lambda
_1),F_2(l;\lambda_2);\theta\bigr)
\nonumber
\\
&{}-\ C\bigl(F_1(k;\lambda_1),F_2(l -
1;\lambda_2);\theta\bigr)
\nonumber
\\
&{}+\ C\bigl(F_1(k - 1;\lambda_1),F_2(l
- 1;\lambda_2);\theta\bigr),\quad k\ge1, l\ge1.\label{EQ:PMF}
\end{align}

Our estimation method is based on the approximation of covariance
$\gamma(\hat\lambda^{\rm CLS}_1,\break\hat\lambda_2^{\rm CLS};\theta)$ by
%
\begin{align}
\label{EQ:COV_APPROX} \gamma^{(M_1,M_2)}\bigl(\hat\lambda^{\rm CLS}_1,
\hat\lambda_2^{\rm
CLS};\theta\bigr) &= \sum
_{k=1}^{M_1} \sum_{l=1}^{M_2}
kl \,c\bigl(F_1\bigl(k; \hat{\lambda}_1^{\rm
CLS}
\bigr), F_2\bigl(l;\hat{\lambda}_2^{\rm CLS}
\bigr);\theta\bigr) - \hat{\lambda}_1^{\rm
CLS}\hat{
\lambda}_2^{\rm CLS}.
\end{align}
For example, if the marginals are Poisson with parameters $\lambda
_1=\lambda_2=1$ and their joint distribution is given by the FGM copula\index{FGM copula} in
\eqref{EQ:FGM}, then the covariance $\gamma^{(M_1,M_2)}(1,1;\theta)$
stops changing significantly after setting
$M_1 = M_2 = M = 8$, regardless of the selected dependence parameter
$\theta$. We used this approximation methodology when carrying out a
Monte Carlo simulation in Section \ref{CH:MC_COP}.

For the FGM copula,\index{FGM copula} if we take the derivative of the sum
%
\begin{align}
\label{EQ:CLS_COPMM} S^{(M_1,M_2)}&= \sum_{t=2}^N
\bigl(R^{\rm CLS}_{1,t}R^{\rm CLS}_{2,t}-
\gamma^{(M_1,M_2)}\bigl(\hat\lambda^{\rm CLS}_1,\hat
\lambda_2^{\rm
CLS};\theta\bigr) \bigr)^2,
\end{align}
equate it to zero and use equation \eqref{EQ:COV_APPROX}, we get
%
\begin{align}
\label{THETA:CLS_THETA_EST}
\hat{\theta}^{\rm FGM}
 \,{=}\, \frac{\sum_{t=2}^N (X_{1,t} - \hat{\alpha
}^{\rm CLS}_1 X_{1,t-1} - \hat{\lambda}^{\rm CLS}_1)(X_{2,t} - \hat
{\alpha}^{\rm CLS}_2 X_{2,t-1} - \hat{\lambda}^{\rm CLS}_2)}
{(N\!-\!1)\sum_{k=1}^{M_1} k( F_{1,k}\overline{F}_{1,k}\,{-}\,F_{1,k-1}\overline
{F}_{1,k-1})\sum_{l=1}^{M_2} l
(F_{2,l}\overline{F}_{2,l}{-}F_{2,l-1}\overline{F}_{2,l-1})},
\end{align}
where $F_{j,k}: = F_j(k;\hat\lambda^{\rm CLS}_j)$, $\overline{F}_{j,k}
:= 1 - F_{j,k}$, $j=1,2$. The derivation of equation \eqref
{THETA:CLS_THETA_EST} is straightforward and thus omitted.

Depending on the selected copula family,
calculation of \eqref{EQ:PMF} to
get the analytical expression of the estimator $\hat{\theta}$ may be
difficult. However, we can use the function \verb|optim| in
the R statistical software to
minimize \eqref{EQ:CLS_COP}. For other
cases,
where the marginal distribution\index{marginal distributions} has parameters other than expected
value $\lambda_j$, equation \eqref{EQ:CLS_COP} would need to be
minimized by those additional parameters.
For example, in the case of negative binomial\index{negative binomial} marginals with
corresponding mean $\lambda_j$ and variance $\sigma_j^2$, i.e.\
when
\begin{align*}
\mathbb{P} (R_{j,t}=k) &= \frac{\varGamma (k+\frac{\lambda_j^2}{\sigma
^2_j-\lambda_j} )}{\varGamma (\frac{\lambda_j^2}{\sigma^2_j-\lambda
_j} ) k!} \biggl(
\frac{\lambda_j}{\sigma_j^2} \biggr)^{\frac{\lambda
_j^2}{\sigma_j^2-\lambda_j}} \biggl(\frac{\sigma^2_j-\lambda_j}{\sigma
_j^2}
\biggr)^k, \quad  k=0,1,\dots,\ j=1,2,
\end{align*}
%
the additional parameters are $\sigma_1^2,\sigma_2^2$, and
the minimization problem becomes
\begin{align*}
S^{(M_1,M_2)}& \longrightarrow\min_{\sigma^2_1,\sigma^2_2, \theta}.
\end{align*}

\subsection{Conditional maximum likelihood estimation}
\noindent
BINAR\index{BINAR}(1) models can be estimated via conditional maximum likelihood\index{conditional maximum likelihood (CML)}
(CML) (see \cite{Pedeli:2011} and \cite{Karlis:2013}).
The conditional distribution of the BINAR\index{BINAR}(1) process is:
\begin{align*}
\mathbb{P}&(X_{1,t} = x_{1,t}, X_{2,t} =
x_{2,t} | X_{1,t-1} = x_{1,t-1}, X_{2, t-1}
= x_{2,t-1})
\\
&= \mathbb{P}(\alpha_1 \circ x_{1,t-1} +
R_{1,t} = x_{1,t}, \alpha_2 \circ
x_{2,t-1} + R_{2,t} = x_{2,t})
\\
&= \sum_{k = 0}^{x_{1,t}} \sum
_{l = 0}^{x_{2,t}}\mathbb{P}(\alpha_1
\circ x_{1,t-1} \,{=}\, k) \mathbb{P}(\alpha_2 \circ
x_{2,t-1}\,{=}\,l)\mathbb {P}(R_{1,t}\,{=}\, x_{1,t} - k,
R_{2,t}\,{=}\, x_{2,t} - l).
\end{align*}
Here, $\alpha_j \circ x$ is the sum of $x$ independent Bernoulli
trials. Hence,
\begin{align*}
\mathbb{P}(\alpha_j \circ x_{j,t-1} = k) =
{x_{j,t-1}\choose k}\alpha_j^k (1-
\alpha_j)^{x_{j,t-1}-k}, \ \ k=0,\dots, x_{j,t-1},\
j=1,2.
\end{align*}

In the case of copula-based BINAR(1) model with Poisson marginals,
\begin{align*}
\mathbb{P}(R_{1,t} = x_{1,t} - k, R_{2,t} =
x_{2,t} - l) &= c\bigl(F_1(x_{1,t} - k,
\lambda_1), F_2(x_{2,t} - l,
\lambda_2); \theta\bigr).
\end{align*}
Thus, we obtain
\begin{align*}
\mathbb{P}&(X_{1,t} = x_{1,t}, X_{2,t} =
x_{2,t} | X_{1,t-1} = x_{1,t-1}, X_{2, t-1}
= x_{2,t-1})
\\
&= \sum_{k = 0}^{x_{1,t}} \sum
_{l = 0}^{x_{2,t}} {x_{1,t-1}\choose
k}
\alpha_1^k (1-\alpha_1)^{x_{1,t-1}-k}
{x_{2,t-1}\choose l}\alpha_2^l (1-
\alpha_2)^{x_{2,t-1}-l}
\\
& \ \ \times c\bigl(F_1(x_{1,t} - k,
\lambda_1), F_2(x_{2,t} - l,
\lambda_2); \theta\bigr)
\end{align*}
and the log conditional likelihood function, for estimating the
marginal distribution\index{marginal distributions} parameters $\lambda_1, \lambda_2$,
the probabilities of the Bernoulli trial successes $\alpha_1, \alpha_2$
and the dependence parameter $\theta$, is
\begin{align*}
\ell(\alpha_1, \alpha_2, \lambda_1,
\lambda_2, \theta) = \sum_{t =
2}^N
\log\mathbb{P}(&X_{1,t} = x_{1,t}, X_{2,t} =
x_{2,t} | X_{1,t-1} = x_{1,t-1},\\
&{}X_{2, t-1} = x_{2,t-1})
\end{align*}
for some initial values $x_{1,1}$ and $x_{2,1}$. In order to estimate
the unknown parameters we maximize the log conditional likelihood:
%
\begin{align}
\label{EQ:CML} \ell(\alpha_1, \alpha_2,
\lambda_1, \lambda_2, \theta) \longrightarrow \max
_{\alpha_1, \alpha_2, \lambda_1, \lambda_2, \theta}.
\end{align}
Numerical maximization\index{numerical maximization} is straightforward with the {\tt optim} function
in the R statistical software.

As 
for the CLS estimator\index{CLS estimators},
in other 
cases, where the marginal distribution\index{marginal distributions} has parameters other than $\lambda_j$,
equation \eqref{EQ:CML} would need to be maximized by those additional
parameters.
The CML estimator is asymptotically normally distributed under standard
regularity conditions and its variance matrix is the inverse of the
Fisher information matrix \cite{Pedeli:2011}.

\subsection{Two-step estimation based on {\rm CLS} and {\rm CML}}

Depending on the range of attainable values of the parameters and the
sample size, CML maximization might take some
time to compute. On the other hand, since CLS estimators\index{CLS estimators} of $\alpha_j$
and $\lambda_j$ are easily derived (compared
to the CLS estimator\index{CLS estimators} of $\theta$, which depends on the copula pmf\index{copulas ! pmf} form
and needs to be numerically maximized),
we can substitute the parameters of the marginal distributions\index{marginal distributions} in
eq.~\eqref{EQ:CML} with CLS estimates\index{CLS estimates} from
equations \eqref{EQ:CLS_ALPHA} and \eqref{EQ:CLS_LAMBDA}. Then we will
only need to maximize $\ell$ with respect to
a single dependence parameter $\theta$ for the Poisson marginal
distribution\index{marginal distributions} case.

Summarizing, the two-step approach to estimating unknown parameters
is to find
\begin{align*}
\bigl(\hat\alpha^{\rm CLS}_j,\hat\lambda_j^{\rm CLS}
\bigr) &= \arg\min Q_j(\alpha_j,
\lambda_j), \quad j=1,2,
\end{align*}
and to take these values as given in the second step:
\begin{align*}
\hat\theta^{\rm CML} &= \arg\max\ell\bigl(\hat\alpha^{\rm CLS}_1,
\hat \alpha_2^{\rm CLS},\hat\lambda^{\rm CLS}_1,
\hat\lambda_2^{\rm CLS}, \theta\bigr).
\end{align*}
For other cases of marginal distribution\index{marginal distributions}, any additional parameters, other
than $\alpha_j$ and $\lambda_j$ would be estimated in the second step.

\subsection{Comparison of estimation methods via Monte Carlo simulation}
\label{CH:MC_COP}

We carried out a Monte Carlo simulation 1000 times to test the
estimation methods with sample size 50 and 500.
The generated model was
a BINAR\index{BINAR}(1) with innovations
joined by either the FGM, Frank or Clayton
copula\index{Clayton copula} with Poisson marginal distributions\index{marginal distributions},
as well as with
marginal distributions\index{marginal distributions}
from different families: one is a Poisson distribution
and the other is a negative binomial one.\index{negative binomial distribution}
Note that for the
two-step method only the estimates of $\theta$ and $\sigma_2^2$ are
included because estimated values of $\alpha_1^{\rm CLS}, \alpha_2^{\rm
CLS}, \lambda_1^{\rm CLS}, \lambda_2^{\rm CLS}$ are used in order to
estimate the remaining parameters via CML.

%
\begin{table}[t]
\caption{Monte Carlo simulation results for a BINAR(1) model with
Poisson innovations linked by the FGM, Frank or Clayton copula}
\label{MC_BINAR}
\tabcolsep=2.5pt
\begin{tabular*}{\textwidth}{@{\extracolsep{\fill}}c|c|c|c|c|c|c|c|c|c@{}}
\hline
\multirow{2}{*}{Copula} &
\multirow{2}{8mm}{\centering Sample size} &
\multirow{2}{*}{Parameter} &
\multirow{2}{8mm}{\centering True value} &
\multicolumn{2}{c|}{CLS} &
\multicolumn{2}{c|}{CML} &
\multicolumn {2}{c}{Two-Step}                    \\\cline{5-10}
&& & & MSE                & Bias     & \multicolumn{1}{c|}{MSE}      & Bias      & \multicolumn{1}{c|}{MSE}              & Bias     \\
\hline
\multirow{10}{*}{FGM}     & \multirow{5}{*}{$N = 50$}          & $\alpha_1$                 & 0.6   & 0.01874                  & $-$0.05823 & \multicolumn{1}{l|}{0.00887}  & $-$0.01789  & \multicolumn{1}{c|}{--}                & --        \\
                          &           & $\alpha_2$                 & 0.4   & 0.02033                  & $-$0.05223 & \multicolumn {1}{l|}{0.01639} & $-$0.02751  & \multicolumn{1}{c|}{--}                & --        \\
                          &   & $\lambda_1$                & 1     & 0.12983                  & 0.13325  & \multicolumn {1}{l|}{0.06514} & 0.03366   & \multicolumn{1}{c|}{--}                & --        \\
                          &           & $\lambda_2$                & 2     & 0.25625                  & 0.16029  & \multicolumn{1}{l|}{0.19939}  & 0.07597   & \multicolumn{1}{c|}{--}                & --        \\
                          &           & $\theta$                   & $-$0.5  & \textbf{0.29789}         & 0.12568  & \multicolumn {1}{l|}{0.33840} & 0.07568   & \multicolumn{1}{l|}{0.3311}           & 0.0876   \\
                          \cline{2-10}
                          & \multirow{5}{*}{$N = 500$}          & $\alpha_1$                 & 0.6   & 0.00147                  & $-$0.00432 & \multicolumn {1}{l|}{0.00073} & $-$0.00122  & \multicolumn{1}{c|}{--}                & --        \\
                          &           & $\alpha_2$                 & 0.4   & 0.00184                  & $-$0.00505 & \multicolumn {1}{l|}{0.00129} & $-$0.00157  & \multicolumn{1}{c|}{--}                & --        \\
                          &  & $\lambda_1$                & 1     & 0.01012                  & 0.00968  & \multicolumn {1}{l|}{0.00556} & 0.00215   & \multicolumn{1}{c|}{--}                & --        \\
                          &           & $\lambda_2$                & 2     & 0.02413                  & 0.01843  & \multicolumn{1}{l|}{0.01763}  & 0.00678   & \multicolumn{1}{c|}{--}                & --        \\
                          &           & $\theta$                   & $-$0.5  & 0.04679                  & 0.00668  & \multicolumn{1}{l|}{0.04271}  & $-$0.00700  & \multicolumn{1}{l|}{\textbf{0.04265}} & $-$0.00443 \\\hline\hline
\multirow{10}{*}{Frank}   &\multirow{5}{*}{$N = 50$}           & $\alpha_1$                 & 0.6   & 0.02023                  & $-$0.06039 & \multicolumn{1}{l|}{0.00950}  & $-$0.01965  & \multicolumn{1}{c|}{--}                & --        \\
                          &           & $\alpha_2$                 & 0.4   & 0.02005                  & $-$0.05251 & \multicolumn {1}{l|}{0.01630} & $-$0.02858  & \multicolumn{1}{c|}{--}                & --        \\
                          & & $\lambda_1$                & 1     & 0.13562                  & 0.13536  & \multicolumn {1}{l|}{0.06740} & 0.03625   & \multicolumn{1}{c|}{--}                & --        \\
                          &           & $\lambda_2$                & 2     & 0.25687                  & 0.16392  & \multicolumn{1}{l|}{0.19975}  & 0.08291   & \multicolumn{1}{c|}{--}                & --        \\
                          &           & $\theta$                   & $-$1    & \textbf{1.83454}         & 0.12394  & \multicolumn {1}{l|}{2.05786} & 0.00860   & \multicolumn{1}{l|}{1.97515}          & 0.04216  \\\cline{2-10}
                          & \multirow{5}{*}{$N = 500$}          & $\alpha_1$                 & 0.6   & 0.00153                  & $-$0.00595 & \multicolumn {1}{l|}{0.00075} & $-$0.00249  & \multicolumn{1}{c|}{--}                & --        \\
                          &           & $\alpha_2$                 & 0.4   & 0.00181                  & $-$0.00582 & \multicolumn {1}{l|}{0.00129} & $-$0.00132  & \multicolumn{1}{c|}{--}                & --        \\
                          & & $\lambda_1$                & 1     & 0.01033                  & 0.01269  & \multicolumn {1}{l|}{0.00550} & 0.00421   & \multicolumn{1}{c|}{--}                & --        \\
                          &           & $\lambda_2$                & 2     & 0.02442                  & 0.02129  & \multicolumn{1}{l|}{0.01785}  & 0.00629   & \multicolumn{1}{c|}{--}                & --        \\
                          &           & $\theta$                   & $-$1    & 0.22084                  & 0.01746  & \multicolumn{1}{l|}{0.20138}  & $-$0.01779  & \multicolumn{1}{l|}{\textbf{0.20070}} & $-$0.01342 \\\hline\hline
\multirow{10}{*}{Clayton} & \multirow{5}{*}{$N = 50$}          & $\alpha_1$                 & 0.6   & 0.01826                  & $-$0.05489 & \multicolumn{1}{l|}{0.00799}  & $-$0.013295 & \multicolumn{1}{c|}{--}                & --        \\
                          &           & $\alpha_2$                 & 0.4   & 0.01976                  & $-$0.05057 & \multicolumn {1}{l|}{0.01585} & $-$0.02427  & \multicolumn{1}{c|}{--}                & --        \\
                          & & $\lambda_1$                & 1     & 0.12679                  & 0.12104  & \multicolumn {1}{l|}{0.06080} & 0.01743   & \multicolumn{1}{c|}{--}                & --        \\
                          &           & $\lambda_2$                & 2     & 0.25725                  & 0.15704  & \multicolumn{1}{l|}{0.19934}  & 0.06499   & \multicolumn{1}{c|}{--}                & --        \\
                          &           & $\theta$                   & 1     & 0.71845                  & 0.02621  & \multicolumn{1}{l|}{0.72581}  & 0.22628   & \multicolumn{1}{l|}{\textbf{0.62372}} & 0.13283  \\\cline{2-10}
                          & \multirow{5}{*}{$N = 500$}          & $\alpha_1$                 & 0.6   & 0.00146                  & $-$0.00518 & \multicolumn {1}{l|}{0.00070} & 0.00016   & \multicolumn{1}{c|}{--}                & --        \\
                          &           & $\alpha_2$                 & 0.4   & 0.00189                  & $-$0.00350 & \multicolumn {1}{l|}{0.00120} & $-$0.00049  & \multicolumn{1}{c|}{--}                & --        \\
                          & & $\lambda_1$                & 1     & 0.00973                  & 0.01137  & \multicolumn {1}{l|}{0.00513} & $-$0.00150  & \multicolumn{1}{c|}{--}                & --        \\
                          &           & $\lambda_2$                & 2     & 0.02447                  & 0.01113  & \multicolumn{1}{l|}{0.01707}  & 0.00065   & \multicolumn{1}{c|}{--}                & --        \\
                          &           & $\theta$                   & 1     & 0.11578                  & 0.03556  & \multicolumn{1}{l|}{0.05864}  & 0.04250   & \multicolumn{1}{l|}{\textbf{0.03199}} & $-$0.01342 \\\hline
\end{tabular*}
%
\end{table}

The results for the Poisson marginal distribution\index{marginal distributions} case are provided in
Table~\ref{MC_BINAR}. The results for the case when one innovation
follows a Poisson distribution and the other follows a negative binomial
one\index{negative binomial distribution} are provided in Table~\ref{MC_BINAR_2}. The lowest MSE
values of $\widehat{\theta}$ are highlighted in bold. It is worth
noting that CML estimation via numerical maximization\index{numerical maximization} depends heavily
on the initial parameter values. If the initial values are selected too
low or too high from the actual value, then the global maximum may not
be found. In order to overcome this, we have selected the starting
values equal to the CLS parameter estimates.

As can be seen in Table~\ref{MC_BINAR}, the estimated values of $\alpha
_j$ and $\lambda_j$, $j=1,2$, have a smaller bias and MSE when
parameters are estimated via CML. On the other hand, estimation of
$\theta$ via CLS exhibits a smaller MSE in the Frank copula\index{Frank copula} case for
smaller samples. For larger samples, the estimates of $\theta$ via the
Two-step estimation method are very close to the CML estimates in terms
of MSE and bias, and are closer to the true parameter values
than the CLS estimates.\index{CLS estimates} Furthermore, since in the Two-step estimation
numerical maximization\index{numerical maximization} is only carried out via a single parameter
$\theta$, the initial parameter values have less
effect on the
numerical maximization.\index{numerical maximization}

%
\begin{table}[t!]
\caption{Monte Carlo simulation results for a BINAR(1) model with one
innovation following a Poisson distribution and the other -- a negative
binomial one,
where both innovations are linked by the FGM,
Frank or Clayton copula}
\label{MC_BINAR_2}
\tabcolsep=2pt
\begin{tabular*}{\textwidth}{@{\extracolsep{\fill}}c|c|c|c|c|c|c|c|c|c@{}}
\hline
\multirow{2}{*}{Copula}   &
\multirow{2}{8mm}{\centering Sample size}   &
\multirow{2}{*}{Parameter} &
\multirow{2}{8mm}{\centering True value }   &
\multicolumn{2}{c|}{CLS}              &
\multicolumn{2}{c|}{CML}                            &
\multicolumn {2}{c}{Two-Step}                      \\
\cline{5-10}
&
&
&
& MSE
& Bias
& \multicolumn{1}{c|}{MSE}
& Bias       & \multicolumn{1}{c|}{MSE}              & Bias       \\\hline
\multirow{12}{*}{FGM}     & \multirow{6}{*}{$N = 50$}  & $\alpha_1$                 & 0.6    & 0.01895                  & $-$0.05858 & \multicolumn{1}{l|}{0.00845}           & $-$0.01513 & \multicolumn{1}{c|}{--}                & --          \\
                          &                            & $\alpha_2$                 & 0.4    & 0.01936                  & $-$0.04902 & \multicolumn {1}{l|}{0.00767}          & $-$0.01953 & \multicolumn{1}{c|}{--}                & --          \\
                          &                            & $\lambda_1$                & 1      & 0.12940                  & 0.12812    & \multicolumn{1}{l|}{0.05424}           & 0.01879    & \multicolumn{1}{c|}{--}                & --          \\
                          &                            & $\lambda_2$                & 2      & 0.39724                  & 0.15151    & \multicolumn{1}{l|}{0.24138}           & 0.04833    & \multicolumn{1}{c|}{--}                & --          \\
                          &                            & $\theta$                   & $-$0.5 & 0.31467                  & 0.14070    & \multicolumn{1}{l|}{\textbf {0.29415}} & 0.06674    & \multicolumn{1}{l|}{0.29949}          & 0.09693    \\
                          &                            & $\sigma_2^2$               & 9      & 27.87327                 & 1.15731    & \multicolumn {1}{l|}{15.12863}         & $-$0.14888 & \multicolumn{1}{l|}{21.68229}         & 0.72326    \\\cline{2-10}
                          & \multirow{6}{*}{$N = 500$} & $\alpha_1$                 & 0.6    & 0.00156                  & $-$0.00695 & \multicolumn{1}{l|}{0.00076}           & $-$0.00153 & \multicolumn{1}{c|}{--}                & --          \\
                          &                            & $\alpha_2$                 & 0.4    & 0.00194                  & $-$0.00373 & \multicolumn {1}{l|}{0.00053}          & 0.00016    & \multicolumn{1}{c|}{--}                & --          \\
                          &                            & $\lambda_1$                & 1      & 0.01041                  & 0.01201    & \multicolumn{1}{l|}{0.00543}           & 0.00290    & \multicolumn{1}{c|}{--}                & --          \\
                          &                            & $\lambda_2$                & 2      & 0.03882                  & 0.01843    & \multicolumn{1}{l|}{0.02362}           & $-$0.00057 & \multicolumn{1}{c|}{--}                & --          \\
                          &                            & $\theta$                   & $-$0.5 & 0.06670                  & $-$0.02014 & \multicolumn{1}{l|}{\textbf {0.04298}} & $-$0.00268 & \multicolumn{1}{l|}{0.04313}          & 0.00562    \\
                          &                            & $\sigma_2^2$               & 9      & 6.24237                  & $-$1.99232 & \multicolumn {1}{l|}{1.81265}          & 0.00611    & \multicolumn{1}{l|}{1.85222}          & $-$0.03506 \\\hline\hline
\multirow{12}{*}{Frank}   & \multirow{6}{*}{$N = 50$}  & $\alpha_1$                 & 0.6    & 0.02049                  & $-$0.06064 & \multicolumn{1}{l|}{0.00912}           & $-$0.01594 & \multicolumn{1}{c|}{--}                & --          \\
                          &                            & $\alpha_2$                 & 0.4    & 0.01951                  & $-$0.04936 & \multicolumn {1}{l|}{0.00772}          & $-$0.02070 & \multicolumn{1}{c|}{--}                & --          \\
                          &                            & $\lambda_1$                & 1      & 0.13769                  & 0.13467    & \multicolumn{1}{l|}{0.05748}           & 0.02280    & \multicolumn{1}{c|}{--}                & --          \\
                          &                            & $\lambda_2$                & 2      & 0.40626                  & 0.15408    & \multicolumn{1}{l|}{0.23717}           & 0.05534    & \multicolumn{1}{c|}{--}                & --          \\
                          &                            & $\theta$                   & $-$1   & 1.81788                  & 0.12516    & \multicolumn{1}{l|}{1.75638}           & $-$0.01239   & \multicolumn{1}{l|}{\textbf{1.68019}} & 0.06211    \\
                          &                            & $\sigma_2^2$               & 9      & 25.10400                 & 0.49423    & \multicolumn {1}{l|}{14.86812}         & $-$0.10034 & \multicolumn{1}{l|}{21.92090}         & 0.74026    \\\cline{2-10}
                          & \multirow{6}{*}{$N = 500$} & $\alpha_1$                 & 0.6    & 0.00161                  & $-$0.00702 & \multicolumn{1}{l|}{0.00075}           & $-$0.00239 & \multicolumn{1}{c|}{--}                & --          \\
                          &                            & $\alpha_2$                 & 0.4    & 0.00187                  & $-$0.00364 & \multicolumn {1}{l|}{0.00050}          & $-$0.00046 & \multicolumn{1}{c|}{--}                & --          \\
                          &                            & $\lambda_1$                & 1      & 0.01093                  & 0.01652    & \multicolumn{1}{l|}{0.00562}           & 0.00501    & \multicolumn{1}{c|}{--}                & --          \\
                          &                            & $\lambda_2$                & 2      & 0.03728                  & 0.01217    & \multicolumn{1}{l|}{0.02335}           & 0.00203    & \multicolumn{1}{c|}{--}                & --          \\
                          &                            & $\theta$                   & $-$1   & 0.31942                  & $-$0.05593 & \multicolumn{1}{l|}{\textbf {0.18960}} & $-$0.01481 & \multicolumn{1}{l|}{0.1902}           & $-$0.0079  \\
                          &                            & $\sigma_2^2$               & 9      & 4.82620                  & $-$1.75765 & \multicolumn {1}{l|}{1.83082}          & 0.02144    & \multicolumn{1}{l|}{1.85852}          & $-$0.02690 \\\hline\hline
\multirow{12}{*}{Clayton} & \multirow{6}{*}{$N = 50$}  & $\alpha_1$                 & 0.6    & 0.01987                  & $-$0.06159 & \multicolumn{1}{l|}{0.00903}           & $-$0.01671 & \multicolumn{1}{c|}{--}                & --          \\
                          &                            & $\alpha_2$                 & 0.4    & 0.01879                  & $-$0.04928 & \multicolumn {1}{l|}{0.00632}          & $-$0.01644 & \multicolumn{1}{c|}{--}                & --          \\
                          &                            & $\lambda_1$                & 1      & 0.13479                  & 0.14072    & \multicolumn{1}{l|}{0.06096}           & 0.03052    & \multicolumn{1}{c|}{--}                & --          \\
                          &                            & $\lambda_2$                & 2      & 0.40675                  & 0.14807    & \multicolumn{1}{l|}{0.23171}           & 0.02871    & \multicolumn{1}{c|}{--}                & --          \\
                          &                            & $\theta$                   & 1      & 0.78497                  & 0.07464    & \multicolumn{1}{l|}{0.67837}           & 0.21235    & \multicolumn{1}{l|}{\textbf{0.57454}} & 0.10972    \\
                          &                            & $\sigma_2^2$               & 9      & 24.40051                 & 0.17321    & \multicolumn {1}{l|}{15.29879}         & $-$0.08379 & \multicolumn{1}{l|}{23.73506}         & 0.73754    \\\cline{2-10}
                          & \multirow{6}{*}{$N = 500$} & $\alpha_1$                 & 0.6    & 0.00153                  & $-$0.00722 & \multicolumn{1}{l|}{0.00075}           & $-$0.00197 & \multicolumn{1}{c|}{--}                & --          \\
                          &                            & $\alpha_2$                 & 0.4    & 0.00196                  & $-$0.00385 & \multicolumn {1}{l|}{0.00047}          & $-$0.00083 & \multicolumn{1}{c|}{--}                & --          \\
                          &                            & $\lambda_1$                & 1      & 0.01036                  & 0.01745    & \multicolumn{1}{l|}{0.00517}           & 0.00409    & \multicolumn{1}{c|}{--}                & --          \\
                          &                            & $\lambda_2$                & 2      & 0.03999                  & 0.01227    & \multicolumn{1}{l|}{0.02304}           & 0.00110    & \multicolumn{1}{c|}{--}                & --          \\
                          &                            & $\theta$                   & 1      & 0.09927                  & 0.04408    & \multicolumn{1}{l|}{\textbf {0.05557}} & 0.03556    & \multicolumn{1}{l|}{0.05559}          & 0.02310    \\
                          &                            & $\sigma_2^2$               & 9      & 2.95995                  & $-$0.68733 & \multicolumn {1}{l|}{1.79836}          & 0.01348    & \multicolumn{1}{l|}{1.87740}          & $-$0.02407 \\\hline
\end{tabular*}\vspace{-6pt}
\end{table}

Table~\ref{MC_BINAR_2} demonstrates the estimation
results when
one innovation 
has a Poisson distribution and the other has a negative binomial
one.\index{negative binomial distribution}
With the inclusion of an additional variance parameter,
the CLS estimation methods exhibit larger MSE and bias
than the CML and Two-step estimation methods,
for both the dependence and variance parameter estimates. Furthermore, the MSE of $\hat{\sigma}_2^2$ is 
smallest when 
the CML estimation method is used. On the other hand, both
the Two-step and CML estimation methods produce similar estimates of $\theta
$ in terms of MSE, regardless of sample size and copula function.\index{copulas ! functions}

We can conclude that it is possible to accurately estimate the dependence
parameter via CML using the CLS estimates\index{CLS estimates} of $\hat{\alpha}_j$ and
$\hat{\lambda}_j$. The resulting $\hat{\theta}$ will be closer to the
actual value of $\theta$ 
than $\hat{\theta}^{\rm CLS}$ and will
not differ much from $\hat{\theta}^{\rm CML}$. Additional inference on
the bias of the estimates can be found in Appendix~\ref{CH:APPENDIX}.

\section{Application to loan default data}\label{CH:APPLICATION}
In this section we estimate a BINAR\index{BINAR}(1) model with the joint innovation
distribution modelled by a copula cdf\index{copulas ! cdf} for empirical data. The data set
consists of loan data\index{loans data} which includes loans that have defaulted and
loans\index{loans} that were repaid without missing any payments (non-defaulted loans). We will analyse
and model the dependence between defaulted and non-defaulted loans as
well as the presence of autocorrelation.

\subsection{Loan default data}
\begin{figure}[t!]
\includegraphics[scale=0.97]{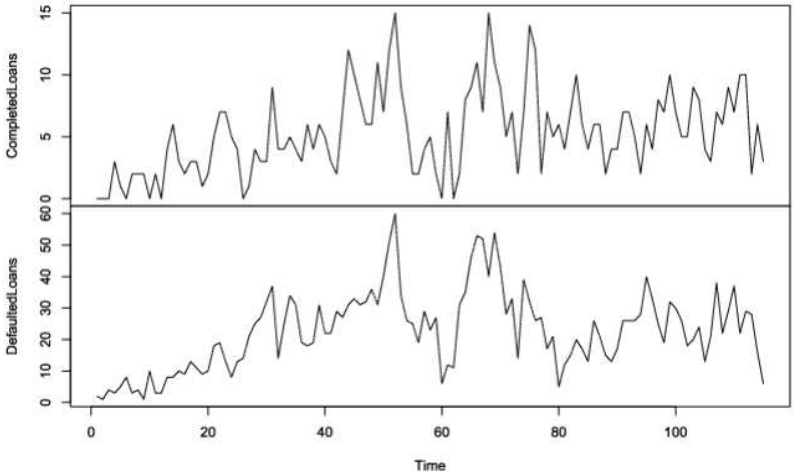}
\caption{Bondora loan data: non-defaulted and defaulted loans by their
issue date}
\label{FIG:LOAN_PLOT1}
\vspace{-7pt}
\end{figure}

The data sample used is from
Bondora, the Estonian peer-to-peer lending company.
In November of 2014 Bondora introduced a loan rating system\index{loans rating system}
which
assigns loans to different groups,
based on their risk level.
There are 
8 groups ranging from the lowest risk group, `AA',
to the highest risk group, `HR'. However, the loan rating
system\index{loans rating system} could not be applied to most older loans\index{older loans} due to a lack of data
needed for Bondora's rating model. 
Although Bondora issues loans\index{Bondora issues loans} in 4
different countries: Estonia, Finland, Slovakia and Spain, we will only
focus on the loans issued in Spain. Since a new rating model indicates
new rules for accepting or rejecting loans,\index{loans} we have selected the data
sample from 21 October 2013, because from that date forward all loans
had a rating assigned to them, to 1 January 2016. The time series are
displayed in Figure \ref{FIG:LOAN_PLOT1}. We are analysing data
consisting of 115 weekly records. 
\begin{itemize}
\item`CompletedLoans' -- the amount of non-defaulted loans\index{loans}
issued per week
which are repaid and have never defaulted (a loan that is 60 or
more days overdue is considered defaulted);
\item`DefaultedLoans' -- the amount of defaulted loans\index{loans}
issued per week.
\end{itemize}
The loan statistics are provided in Table~\ref{TAB:LOAN_TAB1}:
%
\begin{table}[h]
\vspace{-6pt}
\caption{Summary statistics of the weekly data of defaulted
and non-defaulted loans issued in Spain}
\label{TAB:LOAN_TAB1}
%
\begin{tabular}{lrrrr}
\hline
& min & max & mean & variance \\
\hline
DefaultedLoans & 1.00 & 60.00 & 22.60 & 158.66 \\
CompletedLoans & 0.00 & 15.00 & 5.30 & 11.67 \\
\hline
\end{tabular}
\vspace{-6pt}
\end{table}
%
%
\begin{figure}[b!]
\includegraphics[scale=0.97]{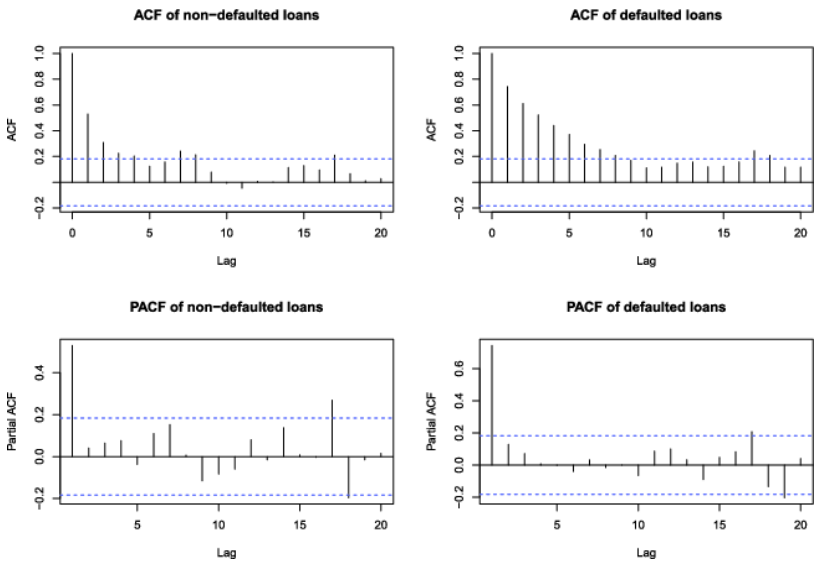}
\caption{AC function and PAC function plots of Bondora loan data}
\label{FIG:LOAN_PLOT_ACF}
\vspace{-7pt}
\end{figure}

The mean, minimum, maximum and variance is higher for defaulted loans\index{defaulted loans}
than for non-defaulted loans\index{loans}. As can be seen from
Figure \ref{FIG:LOAN_PLOT_ACF}, the numbers of defaulted and non-defaulted \index{loans}loans might be correlated since
they both exhibit increase and decrease periods at the same times.

The correlation between the two time series is 0.6684. We also note
that the mean and variance
are lower in the beginning of the time
series. This 
feature could be due to 
various reasons: the effect of
the new loan rating system,\index{loans rating system} which was officially implemented in
December of 2014, the effect of advertising or the fact that the amount
of loans,\index{loans} issued to people living outside of Estonia, increased. The
analysis of the significance of these effects is left for future research.

The sample autocorrelation (AC) function and the partial autocorrelation (PAC)
function are displayed in Figure \ref{FIG:LOAN_PLOT_ACF}. We can
see that the AC function is decaying over time and the PAC function
has a significant first lag which indicates that the non-negative
integer-valued time series could be autocorrelated.

In order to analyse 
if the amount of defaulted loans\index{loans}
depends on the amount of non-defaulted loans\index{loans}
on the same week, we will consider a BINAR\index{BINAR}(1) model with different
copulas for the innovations.\index{innovations} For the marginal distributions\index{marginal distributions} of the
innovations\index{innovations} we will consider the Poisson
distribution as well as the \index{negative binomial distribution}negative binomial one.
Our focus is the estimation of the
dependence parameter, and we will use the Two-step estimation method, based on the Monte Carlo simulation results presented in Section \ref{CH:ESTMATION}.

\subsection{Estimated models}
We estimated a number of BINAR\index{BINAR}(1) models with different distributions
of innovations\index{innovations} which include combinations of:
\begin{itemize}
\item different copula functions: FGM, Frank or Clayton;
\item different combinations of the Poisson and negative binomial
distributions:\index{negative binomial distribution} both marginals are Poisson, both marginals are negative
binomial,\index{negative binomial} or a mix of both.
\end{itemize}

In the first step of the Two-step method, we estimated $\hat{\alpha}_1$
and $\hat{\lambda}_1$ for non-defaulted loans, and $\hat{\alpha}_2$
and $\hat{\lambda}_2$ for defaulted loans\index{defaulted loans} via CLS. The results are
provided in Table~\ref{TAB:APP_TSE_1} with standard errors for the
Poisson case in parenthesis:

%
\begin{table}[t]
\vspace{-6pt}
\caption{Parameter estimates for BINAR(1) model via the Two-step
estimation method: parameter CLS estimates from the first step with
standard errors for the Poisson marginal distribution case in parenthesis}
\label{TAB:APP_TSE_1}
\begin{tabular}{cccc}
\hline
$\hat{\alpha}_1$ & $\hat{\alpha}_2$ & $\hat{\lambda}_1$ & $\hat{\lambda
}_2$ \\\hline
0.53134 & 0.75581 & 2.52174 & 5.58940 \\
(0.08151) & (0.06163) & (0.45012) & (1.41490) \\\hline
\end{tabular}
\vspace{-6pt}
\end{table}

Because the CLS estimation 
of parameters $\alpha_j$ and $\lambda
_j$, $j = 1,2$, does not depend on the selected copula\index{copulas} and the marginal
distribution\index{marginal distributions} family, these parameters will remain the same for each of
the different distribution combinations for innovations.\index{innovations} We can see
that defaulted loans\index{defaulted loans} exhibit a higher degree of autocorrelation 
than non-defaulted loans do,
due to a larger value of $\hat{\alpha}_2$. The innovation mean
parameter for defaulted loans\index{defaulted loans} is also higher,
what
indicates that random shocks have a larger effect on the number of
defaulted loans.\index{defaulted loans}

The 
parameter estimation results from the second-step are provided in
Table~\ref{TAB:APP_TSE_2} with standard errors in parenthesis. $\hat
{\sigma}_1^2$ is the innovation variance estimate of non-defaulted
loans and $\hat{\sigma}_2^2$ is the innovation variance estimate of
defaulted loans.\index{defaulted loans} According to \cite{Pawitan:2001}, the observed Fisher
information is the negative Hessian matrix, evaluated at the maximum likelihood estimator (MLE). The
asymptotic standard errors reported in Table~\ref{TAB:APP_TSE_2} are
derived under the assumption that $\alpha_j$ and $\lambda_j$, $j = 1,2$,
are known, ignoring that the true values are substituted in the second
step with their CLS estimates.\index{CLS estimates}

From the results in Table~\ref{TAB:APP_TSE_2} we see that, according to
the Akaike information criterion (AIC) and log-likelihood values, in most cases the FGM copula\index{FGM copula} most
accurately describes the relationship between the innovations\index{innovations} of
defaulted and non-defaulted loans, with the Frank copula\index{Frank copula} being very close in
terms of the AIC value. The Clayton copula\index{Clayton copula} is the least accurate in describing
the innovation joint distribution, when compared to the FGM and Frank
copula\index{Frank copula} cases, which indicates that defaulted and non-defaulted loans do
not exhibit strong left tail dependence.

Since the summary statistics of the data sample showed that the
variance of the data is larger than the mean, a negative binomial
marginal distribution\index{negative binomial marginal distribution} may provide a better fit. Additionally, because
copulas\index{copulas} can link different marginal distributions,\index{marginal distributions} it is interesting to
see if copulas\index{copulas} with different discrete marginal distributions\index{marginal distributions} would
also improve the model fit. BINAR\index{BINAR}(1) models where non-defaulted loan
innovations\index{innovations} are modelled with negative binomial\index{negative binomial} distributions and defaulted loan
innovations\index{innovations} are modelled with Poisson marginal distributions\index{marginal distributions}, and vice
versa, were estimated. In general, changing one of the marginal
distributions\index{marginal distributions} to a negative binomial\index{negative binomial} provides a better fit in
terms of AIC
than the Poisson marginal distribution\index{marginal distributions} case. However, the
smallest AIC value is achieved when both marginal distributions\index{marginal distributions} are
modelled with negative binomial distributions,\index{negative binomial distribution} linked via the FGM
copula.\index{FGM copula} Furthermore, the estimated innovation variance,
$\hat{\sigma}_2^2$, is much larger
for defaulted loans,\index{defaulted loans}  and this is similar
to what we
observed from the defaulted loan data\index{defaulted loan data} summary statistics.

%
\begin{table}[t]
\caption{Parameter estimates for BINAR(1) model via Two-step estimation
method: parameter CML estimates from the second-step for different
innovation marginal and joint distribution combinations with standard
errors in parenthesis, derived under the assumption that the values
$\hat{\lambda}_j$ and $\hat{\alpha}_j$, $j = 1,2$, from the first step
are true}
\label{TAB:APP_TSE_2}
\begin{tabular*}{\textwidth}{@{\extracolsep{\fill}}l|l|ccc|c|c@{}}
\hline
Marginals                                        & Copula                   & $\hat{\theta}$ & $\hat{\sigma}^2_1$ & $\hat{\sigma }^2_2$ & AIC                                   & Log-likelihood                       \\\hline
\multirow{6}{*}{
{Both Poisson}}                                  & \multirow{2}{*}{FGM}     & 0.89270        & --                  & --                   & \multirow {2}{*}{1763.48096}          & \multirow{2}{*}{$-$880.74048}          \\
                                                 &                          & (0.18671)      &                    &                     &                                       &                                      \\\cline{2-7}
                                                 & \multirow{2}{*}{Frank}   & 2.38484        & --                  & --                   & \multirow{2}{*}{1760.15692}           & \multirow{2}{*}{$-$879.07846}          \\
                                                 &                          & (0.53367)      &                    &                     &                                       &                                      \\\cline{2-7}
                                                 & \multirow{2}{*}{Clayton} & 0.39357        & --                  & --                   & \multirow{2}{*}{1761.12369}           & \multirow{2}{*}{$-$879.56185}          \\
                                                 &                          & (0.11697)      &                    &                     &                                       &                                      \\\hline
\multirow{6}{21mm}{Negative binomial and Poisson}   & \multirow{2}{*}{FGM}     & 1.00000        & 6.46907            & --                   & \multirow {2}{*}{1731.57339}          & \multirow{2}{*}{$-$863.78670}          \\
                                                 &                          & (0.22914)      & (1.01114)          &                     &                                       &                                      \\\cline{2-7}
                                                 & \multirow{2}{*}{Frank}   & 2.14329        & 6.10242            & --                   & \multirow{2}{*}{1731.95241}          & \multirow{2}{*}{$-$863.97620}          \\
                                                 &                          & (0.45100)      & (1.15914)          &                     &                                       &                                      \\\cline{2-7}
                                                 & \multirow{2}{*}{Clayton} & 0.34540        & 5.73731            & --                   & \multirow {2}{*}{1736.47641}          & \multirow{2}{*}{$-$866.23821}          \\
                                                 &                          & (0.12859)      & (0.52831)          &                     &                                       &                                      \\\hline
\multirow{6}{21mm}{Poisson and negative binomial}                              & \multirow{2}{*}{FGM}     & 1.00000        & --                  & 44.83107            & \multirow {2}{*}{1498.29563}          & \multirow{2}{*}{$-$747.14782}          \\
                                                 &                          & (0.26357)      &                    & (7.37423)           &                                       &                                      \\\cline{2-7}
                                                 & \multirow{2}{*}{Frank}   & 2.01486        & --                  & 44.10555            & \multirow {2}{*}{1498.81039}          & \multirow{2}{*}{$-$747.40519}          \\
                                                 &                          & (0.61734)      &                    & (7.33169)           &                                       &                                      \\\cline{2-7}
                                                 & \multirow{2}{*}{Clayton} & 0.38310        & --                  & 43.42739            & \multirow {2}{*}{1503.55388}          & \multirow{2}{*}{$-$749.77694}          \\
                                                 &                          & (0.17376)      &                    & (7.29842)           &                                       &                                      \\\hline
\multirow{6}{21mm}{Both negative binomial} & \multirow {2}{*}{FGM}    & 1.00000        & 6.55810            & 45.36834            & \multirow{2}{*}{\textbf {1466.15418}} & \multirow{2}{*}{\textbf{$-$730.07709}} \\
                                                 &                          & (0.31675)      & (1.24032)          & (7.55217)           &                                       &                                      \\\cline{2-7}
                                                 & \multirow{2}{*}{Frank}   & 2.21356        & 6.58754            & 45.42601            & \multirow {2}{*}{1466.97947}          & \multirow{2}{*}{$-$730.48973}          \\
                                                 &                          & (0.68192)      & (1.26126)          & (7.57743)           &                                       &                                      \\\cline{2-7}
                                                 & \multirow{2}{*}{Clayton} & 0.55939        & 6.64478            & 45.78307            & \multirow {2}{*}{1470.73515}          & \multirow{2}{*}{$-$732.36758}          \\
                                                 &                          & (0.24652)      & (1.25833)          & (7.66324)           &                                       &                                      \\\hline
\end{tabular*}
\end{table}

Overall, both Frank and FGM copulas\index{FGM copula} provide similar fit in terms of
log-likelihood, regardless of the selected marginal distributions.\index{marginal distributions} We
note, however, that for some FGM copula\index{FGM copula} cases, the estimated value of
parameter $\theta$ is equal to the maximal attainable value
1. Based
on copula descriptions\index{copulas ! descriptions} from Section~\ref{CH:COPULA}, the FGM copula\index{FGM copula} is
used to model weak dependence. Given a larger sample size, the Frank
copula\index{Frank copula} might be more appropriate because it can capture a stronger
dependence than
the FGM copula can do.\index{FGM copula}
The negative binomial
marginal distribution\index{negative binomial marginal distribution} case $\hat{\theta} \approx2.21356$ for the Frank
copula\index{Frank copula}
indicates that there is a positive dependence between
defaulted and non-defaulted loans, just as in the FGM copula\index{FGM copula} case.

\section{Conclusions}\label{CH:CONCLUSION}
The analysis via Monte Carlo simulations of different estimation
methods shows that, although the estimates of BINAR\index{BINAR}(1) parameters via
CML has the smallest MSE and bias, estimates of the dependence
parameter has smaller differences of MSE and bias
than for other
estimation methods, indicating that estimations of the dependence
parameter via different 
methods do not exhibit large
differences. While CML estimates exhibit the smallest MSE, their
calculation via numerical optimization relies on the selection of the
initial parameter values. These values can be selected via CLS estimation.

An empirical application of BINAR models for loan data\index{BINAR models for loan data} shows that,
regardless of the selected marginal distributions,\index{marginal distributions} the FGM copula\index{FGM copula}
provides the best model fit in almost all cases. Models with the Frank copula\index{Frank copula} are similar to FGM copula models in terms of AIC values. For some of these cases, the
estimated FGM copula\index{FGM copula} dependence parameter value was equal to the
maximum that can be attained by an FGM copula.\index{FGM copula} In such cases, a larger
sample size could help to determine whether the FGM or Frank copula\index{Frank copula} is more
appropriate to model the dependence between
amounts of defaulted and non-defaulted
loans. 

Although selecting marginal distributions\index{marginal distributions} from different families
(Poisson or negative binomial\index{negative binomial}) provided better models
than those with only Poisson mar\-ginal distributions,\index{marginal distributions} the models with both
marginal distributions\index{marginal distributions} modelled via negative binomial distributions\index{negative binomial distribution}
provide the smallest AIC values which reflects overdispersion in 
amounts of both
defaulted and non-defaulted loans. The FGM copula,\index{FGM copula} which provides the
best model fit, models variables which exhibit weak dependence.
Furthermore, the estimated copula dependence parameter\index{copulas ! dependence parameter} indicates that
the dependence between amounts of defaulted and non-defaulted loans is positive.

Finally, one can apply some other copulas\index{copulas} in order to analyse whether
the loan data\index{loans data} exhibits different forms of dependence from the ones
discussed in this paper. Lastly, the 
approach can be extended by analysing
the presence of structural changes within the data, or checking the
presence of seasonality as well as extending the BINAR\index{BINAR}(1) model with
copula 
joined innovations\index{copulas ! joined innovations} to account for the past values of other time
series rather than only itself.

%
\begin{appendix}
\section{Appendix}\label{CH:APPENDIX}

%
\begin{table}[hb!]
\caption{Standard errors of the bias of the estimated parameters from
the Monte Carlo simulation}
\label{TAB:MC_BIAS_SE}
\tabcolsep=4pt
\begin{tabular*}{\textwidth}{@{\extracolsep{4in minus 4in}}c|c|c|c|c|c|c|c|c|c@{}}
\hline
\multirow{2}{*}{Copula}   & Sample                     & \multirow{2}{*}{Parameter} & True   & \multicolumn{2}{c|}{CLS}           & \multicolumn{2}{c|}{CML}           & \multicolumn {2}{c}{Two-Step}           \\\cline{5-10}
                          & size                       &                            & value  & P-P                      & P-NB    & P-P                      & P-NB    & P-P                           & P-NB    \\\hline
\multirow{12}{*}{FGM}     & \multirow{6}{*}{$N = 50$}  & $\alpha_1$                 & 0.6    & 0.12396                  & 0.12465 & 0.09252                  & 0.09073 & \multicolumn{1}{c|}{--}        & --       \\
                          &                            & $\alpha_2$                 & 0.4    & 0.13274                  & 0.13029 & 0.12510                  & 0.08541 & \multicolumn{1}{c|}{--}        & --       \\
                          &                            & $\lambda_1$                & 1      & 0.33494                  & 0.33631 & 0.25311                  & 0.23225 & \multicolumn{1}{c|}{--}        & --       \\
                          &                            & $\lambda_2$                & 2      & 0.48040                  & 0.61210 & 0.44024                  & 0.48916 & \multicolumn{1}{c|}{--}        & --       \\
                          &                            & $\theta$                   & $-$0.5 & 0.53139                  & 0.54330 & 0.57707                  & 0.53850 & 0.56899                       & 0.53887 \\
                          &                            & $\sigma_2^2$               & 9      & --                        & 5.15368 & --                        & 3.88865 & --                             & 4.60221 \\\cline{2-10}
                          & \multirow{6}{*}{$N = 500$} & $\alpha_1$                 & 0.6    & 0.03813                  & 0.03893 & 0.02706                  & 0.02745 & --                             & --       \\
                          &                            & $\alpha_2$                 & 0.4    & 0.04258                  & 0.04392 & 0.03585                  & 0.02306 & \multicolumn{1}{c|}{--}        & --       \\
                          &                            & $\lambda_1$                & 1      & 0.10018                  & 0.10076 & 0.07455                  & 0.07367 & \multicolumn {1}{c|}{--}       & --       \\
                          &                            & $\lambda_2$                & 2      & 0.15433                  & 0.19676 & 0.13266                  & 0.15377 & \multicolumn {1}{c|}{--}       & --       \\
                          &                            & $\theta$                   & $-$0.5 & 0.21631                  & 0.25760 & 0.20666                  & 0.20741 & 0.20657                       & 0.20770 \\
                          &                            & $\sigma_2^2$               & 9      & --                        & 1.50841 & --                        & 1.34701 & --                             & 1.36119 \\\hline\hline
\multirow{12}{*}{Frank}   & \multirow{6}{*}{$N = 50$}  & $\alpha_1$                 & 0.6    & 0.12882                  & 0.12975 & 0.09552                  & 0.09420 & \multicolumn{1}{c|}{--}        & --       \\
                          &                            & $\alpha_2$                 & 0.4    & 0.13158                  & 0.13073 & 0.12448                  & 0.08543 & \multicolumn{1}{c|}{--}        & --       \\
                          &                            & $\lambda_1$                & 1      & 0.34266                  & 0.34594 & 0.25719                  & 0.23879 & \multicolumn{1}{c|}{--}        & --       \\
                          &                            & $\lambda_2$                & 2      & 0.47982                  & 0.61879 & 0.43939                  & 0.48409 & \multicolumn {1}{c|}{--}       & --       \\
                          &                            & $\theta$                   & $-$1   & 1.34944                  & 1.34314 & 1.43522                  & 1.32589 & 1.40547                       & 1.29538 \\
                          &                            & $\sigma_2^2$               & 9      & --                        & 4.98845 & --                        & 3.85654 & --                             & 4.62540 \\\cline{2-10}
                          & \multirow{6}{*}{$N = 500$} & $\alpha_1$                 & 0.6    & 0.03862                  & 0.03951 & 0.02734                  & 0.02727 & \multicolumn{1}{c|}{--}        & --       \\
                          &                            & $\alpha_2$                 & 0.4    & 0.04212                  & 0.04312 & 0.03591                  & 0.02240 & \multicolumn {1}{c|}{--}       & --       \\
                          &                            & $\lambda_1$                & 1      & 0.10091                  & 0.10329 & 0.07409                  & 0.07481 & \multicolumn {1}{c|}{--}       & --       \\
                          &                            & $\lambda_2$                & 2      & 0.15490                  & 0.19278 & 0.13351                  & 0.15287 & \multicolumn {1}{c|}{--}       & --       \\
                          &                            & $\theta$                   & $-$1   & 0.46985                  & 0.56268 & 0.44862                  & 0.43540 & 0.44802                       & 0.43627 \\
                          &                            & $\sigma_2^2$               & 9      & --                        & 1.31856 & --                        & 1.35359 & --                             & 1.36369 \\\hline\hline
\multirow{12}{*}{Clayton} & \multirow{6}{*}{$N = 50$}  & $\alpha_1$                 & 0.6    & 0.12352                  & 0.12684 & 0.08846                  & 0.09360 & \multicolumn{1}{c|}{--}        & --       \\
                          &                            & $\alpha_2$                 & 0.4    & 0.13123                  & 0.12798 & 0.12361                  & 0.07779 & \multicolumn {1}{c|}{--}       & --       \\
                          &                            & $\lambda_1$                & 1      & 0.33505                  & 0.33926 & 0.24609                  & 0.24514 & \multicolumn {1}{c|}{--}       & --       \\
                          &                            & $\lambda_2$                & 2      & 0.48252                  & 0.62066 & 0.44194                  & 0.48075 & \multicolumn {1}{c|}{--}       & --       \\
                          &                            & $\theta$                   & 1      & 0.84763                  & 0.88328 & 0.82176                  & 0.79618 & 0.77890                       & 0.75037 \\
                          &                            & $\sigma_2^2$               & 9      & --                        & 4.93912 & --                        & 3.91243 & --                             & 4.81812 \\\cline{2-10}
                          & \multirow{6}{*}{$N = 500$} & $\alpha_1$                 & 0.6    & 0.03782                  & 0.03850 & 0.02641                  & 0.02742 & \multicolumn{1}{c|}{--}        & --       \\
                          &                            & $\alpha_2$                 & 0.4    & 0.04337                  & 0.04410 & 0.03468                  & 0.02176 & \multicolumn {1}{c|}{--}       & --       \\
                          &                            & $\lambda_1$                & 1      & 0.09804                  & 0.10033 & 0.07162                  & 0.07180 & \multicolumn {1}{c|}{--}       & --       \\
                          &                            & $\lambda_2$                & 2      & 0.15612                  & 0.19969 & 0.13071                  & 0.15185 & \multicolumn {1}{c|}{--}       & --       \\
                          &                            & $\theta$                   & 1      & 0.33857                  & 0.31212 & 0.23852                  & 0.23316 & 0.23717                       & 0.23476 \\
                          &                            & $\sigma_2^2$               & 9      & --                        & 1.57798 & --                        & 1.34163 & --                             & 1.37066 \\\hline
\end{tabular*}
\end{table}

Let our
Monte Carlo simulation data be
$X_{j,1}^{(i)}, \ldots\, , X_{j,N}^{(i)}$ for simulated sample $i = 1,\ldots\, ,M$
and $j = 1,2$. Let $\eta\in\{\alpha_1, \alpha_2, \lambda_1, \lambda
_2, \theta, \sigma_2^2 \}$ and let $\widehat{\eta}^{(i)}$ be either a
CLS, CML or Two-step estimate of the true parameter value $\eta$ for
the simulated sample $i$.

The mean squared error and the bias are calculated as follows:
\begin{align*}
\text{MSE}( \widehat{\eta}) &= \dfrac{1}{M} \sum
_{i = 1}^M \bigl(\widehat {\eta}^{(i)}
- \eta\bigr)^2,
\\
\text{Bias}( \widehat{\eta}) &= \dfrac{1}{M} \sum
_{i = 1}^M\bigl(\widehat {\eta}^{(i)} -
\eta\bigr).
\end{align*}
Calculating the per-sample bias for each simulated sample $i$ would
also allow us to calculate the sample variance of biases $\text{Bias}(
\widehat{\eta}^{(i)}) = \widehat{\eta}^{(i)} - \eta$, $i = 1,\ldots\, ,M$:
\begin{align*}
\widehat{\mathbb{V} {\rm ar}} \bigl( \text{Bias}( \widehat{\eta}) \bigr) &=
\dfrac{1}{M-1} \sum_{i = 1}^M
\bigl[ \text{Bias}\bigl( \widehat{\eta }^{(i)}\bigr) - \text{Bias}(
\widehat{\eta}) \bigr]^2,
\end{align*}
which we can use to calculate the standard error of the bias.

%
\begin{figure}[t]
\includegraphics{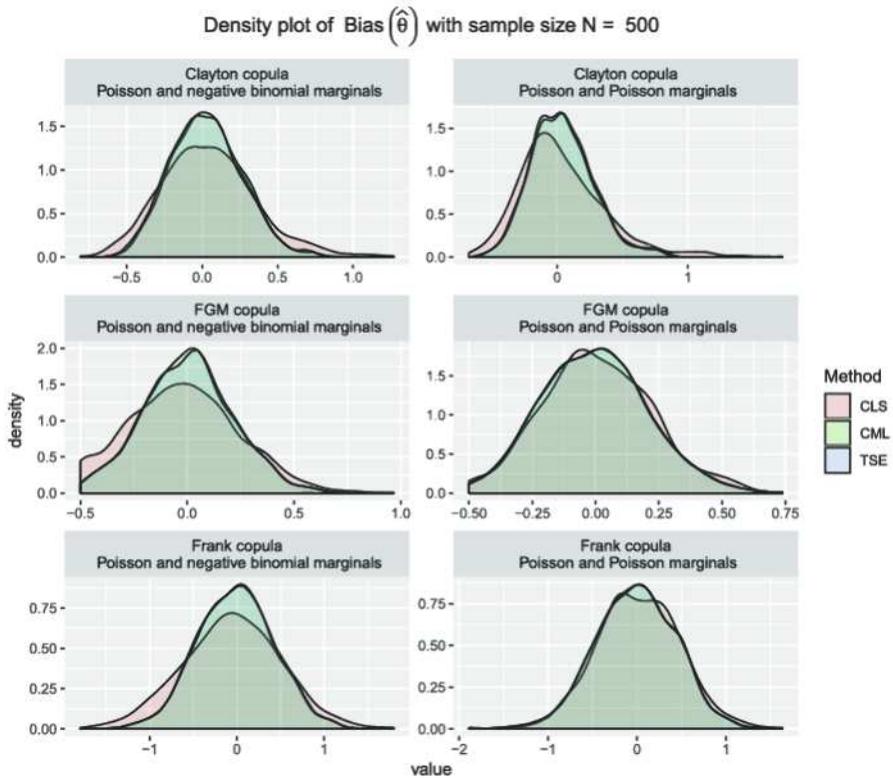}
\caption{Kernel density estimate for 
the bias of the dependence
parameter estimates in the Monte Carlo simulation}
\label{FIG:MC_BIAS_SE}
\vspace{-6pt}
\end{figure}

The standard errors of the parameter bias of the Monte Carlo simulation
are presented in Table~\ref{TAB:MC_BIAS_SE}. The columns labelled `P-P'
indicate the cases where both innovations\index{innovations} have Poisson marginal
distributions,\index{marginal distributions} while columns labelled `P-NB' is for the cases where one
innovation component follows the Poisson distribution and
the other follows a negative
binomial one.\index{negative binomial distribution}
The kernel density estimate for the bias of the
dependence parameter estimate, $\widehat{\theta}$, is presented in
Figure \ref{FIG:MC_BIAS_SE} for the Monte Carlo simulation cases, where
the sample size was 500.

The results in Table~\ref{TAB:MC_BIAS_SE} are in line with the
conclusions presented in Section \ref{CH:MC_COP} -- for $\hat{\alpha
}_j$, $\hat{\lambda}_j$, $j = 1,2$, and $\hat{\sigma}_2^2$ the standard
error of the bias is smaller for CML
than for CLS. On the other
hand, $\hat{\theta}$ has a similar standard error of the bias for CML
and Two-step estimation methods. From Figure \ref{FIG:MC_BIAS_SE} we see
that the CML and Two-step estimates of the dependence parameter $\theta
$ are similar to
each other and have a lower standard error of the
bias 
than the CLS estimate.
\end{appendix}

\begin{acknowledgement}
The authors would like to thank the anonymous referee for his/her
feedback and constructive insights, which helped to improve this paper.
\end{acknowledgement}





\end{document}